\documentclass[10pt]{article}
\usepackage{times}
\usepackage{mathptmx}

\usepackage{graphicx}
\usepackage[fleqn]{amsmath}
\usepackage[square,numbers]{natbib}
\usepackage[letter]{crop}
\usepackage{hyperref}
\usepackage{memhfixc}
\usepackage{amssymb,amsfonts}
\DeclareMathAlphabet{\mathbi}{OML}{cmm}{b}{it} 
\newcommand{\rem}[1]{}

\newcommand{\bel}{\begin{equation}\label}
\newcommand{\ee}{\end{equation}}
\newcommand{\beq}{\begin{eqnarray}\label} 
\newcommand{\eeq}{\end{eqnarray}} 
\newcommand{\bc}{\begin{center}} 
\newcommand{\ec}{\end{center}} 
\newcommand{\ben}{\begin{enumerate}}
\newcommand{\een}{\end{enumerate}}
\newcommand{\bit}{\begin{itemize}}
\newcommand{\eit}{\end{itemize}}
\newtheorem{theorem}{Theorem}

\newtheorem{proposition}{Proposition}

\newcommand{\I}{\int_{\mathcal{V}}}
\newcommand{\bom}{\mbox{\boldmath$\omega$}}
\newcommand{\bu}{\mathbi{u}}

\newcommand{\non}{\nonumber}

\newcommand\fourthirds{\ensuremath{{\scriptstyle\frac{4}{3}}}}

\newcommand\shalf{\ensuremath{{\scriptstyle\frac{1}{2}}}}

\begin{document}

\bc\textbf{\large 
Dynamics of scaled norms of vorticity for the three-dimensional\\
Navier-Stokes and Euler equations}
\ec
\bc
\textbf{J. D. Gibbon\footnote{j.d.gibbon@ic.ac.uk}\\
Department of Mathematics, Imperial College London SW7 2AZ, UK}
\ec

\begin{abstract}
A series of numerical experiments is suggested for the three-dimensional Navier-Stokes and Euler equations on a 
periodic domain based on a set of $L^{2m}$-norms of vorticity $\Omega_{m}$ for $m\geq 1$. These are scaled 
to form the dimensionless sequence $D_{m}= (\varpi_{0}^{-1}\Omega_{m})^{\alpha_{m}}$ where $\varpi_{0}$ 
is a constant frequency and $\alpha_{m} = 2m/(4m-3)$.  A numerically testable Navier-Stokes regularity criterion 
comes from comparing the relative magnitudes of $D_{m}$ and $D_{m+1}$ while another is furnished by imposing 
a critical lower bound on $\int_{0}^{t}D_{m}\,d\tau$. The behaviour of the $D_{m}$ is also important in the Euler 
case in suggesting a method by which possible singular behaviour might also be tested.
\end{abstract}
\bc
To appear in the \textit{Procedia IUTAM} volume of papers \textit{Topological Fluid Dynamics II}.
\ec


\section{\large Introduction}\label{intro}

The challenges that face those concerned with the numerical integration of the three-dimensional incompressible Euler 
and Navier-Stokes equations for a velocity field $\bu(x,\,t)$ on a 3D periodic domain $V = [0,\,L]^{3}_{per}$
\bel{NSdef}
\frac{D\bu}{Dt} = \nu \Delta \bu -\nabla p\qquad\qquad \mbox{div}\,\bu = 0
\ee
($\nu = 0$ for the Euler equations) are also reflected in the challenges faced by analysts in their attempts to understand 
the regularity of these equations. The best known and most effective result in which analysis has guided numerics is the 
Beale-Kato-Majda (BKM) theorem \cite{BKM,MB01}, which says that solutions of the three-dimensional incompressible 
Euler equations are controlled from above by $\int_{0}^{t}\|\bom\|_{\infty}\,d\tau$. If this is finite then no blow-up can 
occur at time $t$. Moreover, any numerical singularity in the vorticity field of the type $\|\bom\|_{\infty}\sim \left(t_{0} 
- t\right)^{-p}$ must have $p\geq 1$ for the singularity not to be a numerical artefact. The BKM criterion has become a 
standard feature in Euler computations\,: see the papers in the special volume \cite{Aussois}. 

This present paper is concerned with regularity criteria that form a consistent framework for the Euler and Navier-Stokes 
equations and which are testable numerically. In both cases it is often not clear when a numerically observed spike 
in the vorticity or strain fields remains finite or is a manifestation of a singularity. It is well known that monitoring the global 
enstrophy, or $H_{1}$-norm $\|\bom\|_{2}$, pointwise in time determines Navier-Stokes regularity \cite{CF88,FMRT}, while 
the monitoring of $\|\bom\|_{\infty}$ likewise determines the fate of Euler solutions. However, the range of $L^{p}$-norms 
between these may be useful. The basic objects are a set of frequencies based on $L^{2m}$-norms of the 
vorticity field $\bom = \mbox{curl}\,\bu$
\bel{Omdef}
\Omega_{m}(t) = \left(L^{-3}\I |\bom|^{2m}dV\right)^{1/2m}\qquad\qquad 1 \leq m \leq \infty\,.
\ee
H\"older's inequality insists that $\Omega_{m} \leq \Omega_{m+1}$. 
The Navier-Stokes and Euler equations are invariant under the re-scaling $x' = \epsilon x$\,; $t' = \epsilon^{2}t$\,; 
$\bu = \epsilon\bu'$\,; $p = \epsilon^{2}p'$. If the domain length $L$ is also re-scaled as  $L'=\varepsilon L$ then 
$\Omega_{m}$  re-scales as $\Omega_{m} = \varepsilon^{2}\Omega_{m}'$ as expected. If, however, $L$ is not 
re-scaled but kept fixed then $\Omega_{m}$ re-scales as 
\bel{scale1}
\Omega_{m}^{\alpha_{m}} = \epsilon \Omega_{m}^{'\,\alpha_{m}}\,,
\ee
where 
\bel{alphadef}
\alpha_{m}=\frac{2m}{4m-3}\,.
\ee
It turns out that this strange scaling is particularly important and provides a motivation for the definition of the set of 
dimensionless quantities
\bel{Dmdef}
D_{m}(t) = \left(\varpi_{\,0}^{-1}\Omega_{m}\right)^{\alpha_{m}}\,,\qquad\qquad 1\leq m \leq \infty\,
\ee
where $\alpha_{1} = 2$ and $\alpha_{\infty} = 1/2$. For the Navier-Stokes equations the frequency $\varpi_{\,0}$ is easily 
defined as $\varpi_{\,0} = \nu L^{-2}$.  The case of the Euler equations is more difficult as there is no obvious material constant 
to replace $\nu$ in the definition of $\varpi_{0}$. A circulation $\Gamma$ has the same dimensions as that of $\nu$ but it must 
be taken around some chosen initial data\,: for instance, in \cite{Kerr93}, initial data was taken to be a pair of anti-parallel vortex 
tubes, in which case $\Gamma$ could be chosen as the circulation around one of these.  For a discussion of the variety of conclusions 
that can be drawn from numerical experiments see \cite{Kerr93,BustKerr08,DHY,Hou08,GHDG08,JDGAussois}. 

No proof exists, as yet, of the existence and uniqueness of solutions of either the 3D Navier-Stokes or Euler equations for arbitrarily 
long times. A time-honoured approach has been to look for minimal assumptions that achieve this result \cite{CF88,FMRT}. In fact, 
much of what is known about solutions of both the 3D Navier-Stokes and Euler equations is encapsulated in the sequence of time 
integrals
\bel{regcon1}
\int_{0}^{t}D_{m}^{2}\,d\tau
\ee
based on the continuum lying between
\bel{sequ1}
D_{1} = \varpi_{\,0}^{-2}L^{-3}\I |\bom|^{2}\,dV~\ldots
~~~~~\to~~~~~
\ldots~D_{\infty} = \left(\varpi_{\,0}^{-1}\|\bom\|_{\infty}\right)^{1/2}\,.
\ee
A well-known time-integral regularity condition for the Navier-Stokes equatons is that the first in the sequence in (\ref{regcon1}) 
should be finite \cite{CF88,FMRT}\,: that is $\int_{0}^{t}D_{1}^{2}\,d\tau < \infty$. In contrast, the boundedness of the last 
in the sequence in (\ref{regcon1}) at $m = \infty$ is exactly the Beale-Kato-Majda criterion $\int_{0}^{t}\|\bom\|_{\infty}\,d\tau 
< \infty$ for the regularity of solutions of the Euler equations \cite{BKM,MB01,Aussois}. 

For three-dimensional Navier-Stokes turbulence, it has to be admitted that arbitrarily imposed regularity assumptions, such as 
$\int_{0}^{t}D_{m}^{2}\,d\tau < \infty$, while mathematically interesting, have little foundation in physics\,: see the discussion 
of this point in \cite{JDGJMP12}. However, what is known, without any assumptions, is that weak solutions (in the sense of Leray 
\cite{Leray34}) obey the time integral \cite{JDGCMS11}
\bel{Dmav1}
\int_{0}^{t}D_{m}\,d\tau \leq c\,\left( t\,Re^{3} + \eta_{1}\right)\,,
\ee
where $\eta_{1}$ is a constant depending upon $D_{m}(0)$. This result plays two roles. In \S\ref{weak} it is shown that it leads to 
a definition of a continuum of inverse length scales $L\lambda_{m}^{-1}$ 
the upper bound on the first of which is the well-known Kolmogorov scale proportional to $Re^{3/4}$. The more general upper
bound is discussed in \S\ref{weak} and is given by $L\lambda_{m}^{-1}\leq c\,Re^{3/2\alpha_{m}}$. Thus the $\lambda_{m}$ 
for $m > 1$ correspond to deeper length scales associated with the higher $L^{2m}$-norms of vorticity implicit within the $D_{m}$. 

In addition, the magnitude of the bounded time integral in (\ref{Dmav1}) is also significant. Let us consider whether the saturation, 
or near saturation, of this time integral plays any role in the regularity question. In the forced case it has been shown in 
\cite{JDGJMP12} that if a critical \textit{lower} bound is imposed on this time integral in terms of the Grashof number $Gr$ then 
this leads to exponential collapse in the $D_{m}(t)$. In fact boundedness from above of any one of the $D_{m}$ also implies 
the boundedness of $D_{1}$ which immediately leads to the existence and uniqueness of solutions.  While it can be argued that
the imposition of this lower bound is physically artificial the result is nevertheless intriguing because it suggests that if the value of 
the integral (\ref{Dmav1}) is sufficiently \textit{large} then solutions are under control, which is surely counter-intuitive. Once it dips 
\textit{below} this critical value then regularity could potentially break down. The importance of this mechanism lies in the role it 
may play in understanding the phenomenon of Navier-Stokes intermittency. This is discussed in \S\ref{body} where the critical 
lower bound is expressed in terms of the more physical Reynolds number $Re$
\bel{Dmav2}
\left( t\,Re^{3\delta_{m}} + \eta_{2}\right) \leq \int_{0}^{t}D_{m}\,d\tau\,,\qquad\qquad 0 \leq \delta_{m} \leq 1\,.
\ee
The range of $\delta_{m}$ is estimated and it is shown that $\delta_{m}\searrow\shalf$ for large $m$, thus allowing considerable 
slack between the critical lower bound in (\ref{Dmav2}) and the upper bound in (\ref{Dmav1}).

The $D_{m}$ are comparatively easy quantities to calculate from a numerical scheme and is thus it is worth exploring whether 
regularity criteria can be gleaned from the relative magnitudes of the $D_{m}$ or their time integrals. This is the task of this 
paper. In \S\ref{Euler} the Euler equations are discussed in these terms where two versions of a numerical experiment are 
suggested for testing singular or non-singular behaviour.  

\section{\large The incompressible $3D$ Euler equations}\label{Euler}

Whether the three-dimensional Euler equations develop a singularity in a finite time still remains an open problem but a variety 
of super-weak solutions have recently been shown to exist \cite{Sh97,deLSz07,deLSz10,Wied11,BardosTiti07,BardosTiti10}. Let 
$\Gamma$ be the circulation around some chosen initial data such that $\varpi_{0} = \Gamma L^{-2}$, as discussed in \S\ref{intro}. 
This defines $\varpi_{0}$ 
within the definition of $D_{m}$. 
\begin{proposition}\label{prop1}
Provided solutions of the three-dimensional Euler equations exist, for $1 \leq m < \infty$ the $D_{m}$ formally satisfy 
the following differential inequality
\bel{propeqn1}
\dot{D}_{m} \leq c_{m}\, \varpi_{\,0} \left(\frac{D_{m+1}}{D_{m}}\right)^{\xi_{m}}D_{m}^{3}\,,
\qquad\qquad
\xi_{m} = \shalf(4m+1)\,.
\ee
\end{proposition}
\textbf{Proof\,:} The time derivative of the $\Omega_{m}$ obeys
\beq{P1}
2mL^{3}\Omega_{m}^{2m-1}\dot{\Omega}_{m}
&=& \frac{d~}{dt}\int |\bom|^{2m}\,dV\non\\
&\leq & 2m\int_{0}^{t}\bom^{2m}|\nabla\bu|\,dV\non\\
& \leq & 2m \left(\int |\bom|^{2m}\,dV\right)^{\frac{1}{2}}
\left(\int |\bom|^{2(m+1)}\,dV\right)^{\frac{m}{2(m+1)}}
\left(\int |\nabla u|^{2(m+1)}\,dV\right)^{\frac{1}{2(m+1)}}
\non\\
&\leq& 2m L^{3}c_{1,m}\,\Omega_{m+1}^{m+1}\,\Omega_{m}^{m}
\eeq
where we have used $\|\nabla u\|_{p} \leq c_{p}\|\omega\|_{p}$, for $1 \leq p < \infty$. 
Thus it transpires that
\bel{P2}
\dot{\Omega}_{m} \leq c_{1,m} \left(\frac{\Omega_{m+1}}{\Omega_{m}}\right)^{m+1}\Omega_{m}^{2}\,,
\qquad\qquad 1 \leq m < \infty\,.
\ee 
Note that the case $m=\infty$ is excluded\,: it was shown in [1] that a logarithmic $H_{3}= \I |\nabla^{3}\bu|^{2}\,dV$ 
factor is needed such that $\|\nabla u\|_{\infty} \leq c\,\|\omega\|_{\infty}\left(1+ \ln H_{3}\right)$.
\par\medskip
We wish to convert inequality (\ref{P2}) to one in terms of $D_{m}$
\beq{P3}
\left(\frac{\Omega_{m+1}}{\Omega_{m}}\right)^{m+1}\Omega_{m} &=& \varpi_{\,0}
\left(\frac{D_{m+1}}{D_{m}}\right)^{\frac{m+1}{\alpha_{m+1}}}
D_{m}^{\left(\frac{m+1}{\alpha_{m+1}} - \frac{m+1}{\alpha_{m}}\right) + \frac{1}{\alpha_{m}}}
= \varpi_{\,0}\left(\frac{D_{m+1}}{D_{m}}\right)^{\shalf(4m+1)}D_{m}^{2}\nonumber
\eeq
having used the fact that 
\bel{P4}
\left(\frac{1}{\alpha_{m+1}} - \frac{1}{\alpha_{m}}\right)\beta_{m} = 2\,,\qquad\qquad \beta_{m} = \fourthirds m(m+1)\,.
\ee
Substitution into (\ref{P2}) completes the proof. \hfill $\blacksquare$
\par\medskip\noindent
There are now at least two interesting routes for the integration of (\ref{propeqn1}). 
\ben
\item Firstly if a finite time singularity is suspected then divide (\ref{propeqn1}) by $D_{m}^{3-\varepsilon}$ 
with $0 \leq \varepsilon < 2$ to obtain
\bel{E6}
[D_{m}(t)]^{2-\varepsilon} \leq \frac{1}{[D_{m}(t_{0})]^{-(2-\varepsilon)} - F_{1,\varepsilon}(t)}
\ee
where
\bel{F1epdef}
F_{1,\varepsilon}(t) = c_{m} (2-\varepsilon)\varpi_{\,0} 
\int_{t_{0}}^{t} \left(\frac{D_{m+1}}{D_{m}}\right)^{\xi_{m}}D_{m}^{\varepsilon}\,d\tau\,,
\qquad\qquad \xi_{m} = \shalf(4m+1)\,.
\ee
For instance, for $\varepsilon = 0$ we have 
\bel{E4}
D_{m}^{2}(t) \leq \frac{1}{[D_{m}(t_{0})]^{-2} - F_{1,0}(t)}
\ee
where
\bel{F1def}
F_{1,0}(t) = 2c_{m} \varpi_{\,0} \int_{t_{0}}^{t} \left(\frac{D_{m+1}}{D_{m}}\right)^{\xi_{m}}\,d\tau\,.
\ee
A singularity in the upper bound of inequality (\ref{E4}) is not necessarily significant. What is more significant 
is whether the solution tracks this singular upper bound. This suggests the following numerical test\,:
\ben
\item Is $F_{1,0}$ linear in $t$?
\item If so, then test whether
\bel{test1}
D_{m}^{2}\left(T_{c,m} - t\right) \to C_{m}\qquad\qquad\mbox{with}\qquad\qquad T_{c,m}\to T_{c}
\ee
uniformly in $m$. If such behaviour occurs it suggests, but does not prove, that the $D_{m}$ may be blowing 
up close to the upper bound. 
\een

\item If exponential or super-exponential growth is  suspected then divide (\ref{propeqn1}) only by 
$D_{m}$ (the case $\varepsilon = 2$) and integrate
\bel{E5}
D_{m} \leq D_{m}(t_{0}) \exp \varpi_{\,0} \int_{t_0}^{t} \left(\frac{D_{m+1}}{D_{m}}\right)^{\xi_{m}}D_{m}^{2}\,d\tau\,.
\ee
The rate of growth of the integral with respect to $t$ is of interest. Does it remain finite for as long as the integrated 
solution remains reliable? 
\een
The companion paper is this volume by Kerr addresses some of these questions \cite{KerrTOD}. 

\section{\large Weak solutions of Navier-Stokes and a range of scales}\label{weak}

Weak solutions are natural for the global enstrophy $\|\bom\|_{2}^{2}$ because of the properties of projection operators. The 
original argument used by Leray \cite{Leray34} gives us the textbook result from his energy inequality \cite{CF88,FMRT}. In terms 
of $D_{1}$ this is 
\bel{Leray1}
\left<D_{1}\right>_{T} \leq c\,Re^{3} + O\left(T^{-1}\right)
\ee
where the time average up time $T$ given by $\left<\cdot\right>_{T}$ is defined by
\bel{timavdef1}
\left<F(\cdot)\right>_{T} = \frac{1}{T}\limsup_{F_{0}}\int_{0}^{T}F(\tau)\,d\tau\,.
\ee
To obtain similar results for  $\|\bom\|_{2m}$ for $m > 1$ looks difficult not only because 
the properties of projection operators do not naturally extend to the higher spaces but also because 
$\|\bom\|_{2m}$ does not appear naturally in an energy inequality. However, these problems 
have been circumvented in \cite{JDGCMS11}, the main result from which will be stated below and 
its very short proof repeated for the benefit of the reader\,:
\begin{theorem}\label{weakthm} For $1 \leq m \leq \infty$, weak solutions obey
\bel{Ombd1}
\left<D_{m}\right>_{T} \leq c\,Re^{3} + O\left(T^{-1}\right)\,,
\ee
where $c$ is a uniform constant.
\end{theorem}
\par\medskip\noindent
\textbf{Proof\,:} The proof is based on a result of Foias, Guillop\'e and Temam \cite{FGT} (their Theorem 3.1) for weak 
solutions. Doering and Foias \cite{DF02} used the square of the averaged velocity $U_{0}^{2} = 
L^{-3}\left<\|\bu\|_{2}^{2}\right>_{T}$ to define the Reynolds number $Re = U_{0}L\nu^{-1}$ which enables us to
convert estimates in $Gr$ to estimates in $Re$.  Thus the result of Foias, Guillop\'e and Temam \cite{FGT} in terms of $Re$
becomes
\bel{FGT1a}
\left< H_{N}^{\frac{1}{2N-1}}\right>_{T} \leq c_{N}L^{-1}\nu^{\frac{2}{2N-1}}Re^{3} + O\left(T^{-1}\right)\,,
\ee
where
\bel{HNdef}
H_{N} = \I \left|\nabla^{N}\bu\right|^{2}\,dV = \I\left|\nabla^{N-1}\bom\right|^{2}\,dV\,,
\ee
and where $H_{1} = \I \left|\nabla\bu\right|^{2}\,dV = \I \left|\bom\right|^{2}\,dV$. Then an interpolation 
between $\|\bom\|_{2m}$ and $\|\bom\|_{2}$ is written as
\bel{FGT2} 
\|\bom\|_{2m} \leq c_{N,m} \|\nabla^{N-1}\bom\|_{2}^{a}\,\|\bom\|_{2}^{1-a}\,,\qquad\qquad  
a = \frac{3(m-1)}{2m(N-1)}\,,
\ee
for $N\geq 3$. $\|\bom\|_{2m}$ is raised to the power $A_{m}$, which is to be determined.
\beq{FGT3}
\left<\|\bom\|_{2m}^{A_{m}}\right>_{T} &\leq& 
c_{N,m}^{A_{m}} \left<\|\nabla^{N-1}\bom\|_{2}^{a A_{m}}
\|\bom\|_{2}^{(1-a) A_{m}}\right>_{T}\non\\
&=& c_{N,m}^{A_{m}} \left<\left(H_{N}^{\frac{1}{2N-1}}\right)^{\shalf a A_{m}(2N-1)}
H_{1}^{\shalf(1-a) A_{m}}\right>_{T}\non\\
&\leq& c_{N,m}^{A_{m}} \left<H_{N}^{\frac{1}{2N-1}}\right>_{T}^{\shalf a A_{m}(2N-1)}
\left<H_{1}^{\frac{(1-a) A_{m}}{2-a A_{m}(2N-1)}}\right>_{T}^{1-\shalf a A_{m}(2N-1)}
\eeq
An explicit upper bound in terms of $Re$ is available only if the exponent of $H_{1}$ within the 
average is unity\,; that is
\bel{FGT4}
\frac{(1-a)A_{m}}{2-a A_{m}(2N-1)} = 1\qquad\Rightarrow\qquad
A_{m} = \frac{2m}{4m-3} = \alpha_{m}
\ee
as desired. Using the estimate in (\ref{FGT1a}), and (\ref{Leray1}) for $\left<H_{1}\right>$, the result 
follows. $c_{N,m}$ can be minimized by choosing $N=3$. $c_{3,m}$ does not blow up even 
when $m=\infty$\,; thus we take the largest value of $c_{3,m}^{\alpha_{m}}$ and call this 
$c$. \hfill $\blacksquare$

\par\medskip
Following the statement of Theorem \ref{weakthm} and motivated by the definition of the Kolmogorov length for $m=1$, 
a continuum of length scales $\lambda_{m}$ can be defined thus\,:
\bel{lamdef1}
\left< D_{m}\right>_{T} : = \left(L\lambda_{m}^{-1}\right)^{2\alpha_{m}}\,,
\ee
in which case (\ref{lamdef1}) becomes
\bel{lamdef2}
L\lambda_{m}^{-1} \leq c\,Re^{3/2\alpha_{m}}\,.
\ee
When $m=1$, $\alpha_{1}=2$, and thus $L\lambda_{1}^{-1} \leq c\,Re^{3/4}$, which is consistent with 
Kolmogorov's statistical theory [22,\,23]. However, the bounds on $\lambda_{m}^{-1}$ become increasingly 
large with increasing $m$ reflecting how the $L^{2m}$-norms can detect finer scale motions.

\section{\large A regularity criterion based on the relative sizes of $D_{m}$ and $D_{m+1}$}\label{log}

Consider two $m$-dependent constants $c_{1,m}$ and $c_{2,m}$ and two 
frequencies $\varpi_{1,m}$ and $\varpi_{2,m}$ defined by
\bel{varpidef}
\varpi_{1,m} = \varpi_{0}\alpha_{m}c_{1,m}^{-1}\qquad\qquad
\varpi_{2,m} = \varpi_{0}\alpha_{m}c_{2,m}\,.
\ee
In \cite{JDGJMP12}, using a standard contradiction strategy on a finite interval of existence and uniquness $[0,\,T^{*})$, 
it was shown that for the decaying Navier-Stokes equations the $D_{m}$ obey the following theorem in which the dot 
represents differentiation with respect to time\,:
\begin{theorem}\label{Dmthm}
For $1 \leq m < \infty$ on $[0,\,T^{*})$ the $D_{m}(t)$ satisfy the set of inequalities 
\bel{Dminequal}
\dot{D}_{m} \leq D_{m}^{3}
\left\{-  \varpi_{1,m}\left(\frac{D_{m+1}}{D_{m}}\right)^{\rho_{m}} + \varpi_{2,m}\right\}\,,
\ee
where $\rho_{m} = \frac{2}{3} m(4m+1)$.  In the forced case there is an additive term $\varpi_{3,m}Re^{2}D_{m}$.
\end{theorem}
\par\medskip\noindent
The obvious conclusion is that solutions come under control pointwise in $t$ provided 
\bel{Dmcon1}
D_{m+1}(t) \geq c_{\rho_{m}} D_{m}(t)
\ee
where 
\bel{Dmcon2}
c_{\rho_{m}} =  \left[c_{1,m}c_{2,m}\right]^{1/\rho_{m}}
\ee
This is a numerically testable criterion\,: if (\ref{Dmcon1}) holds then the $D_{m}$ must decay in time.
It was shown in \cite{JDGJMP12} that this can be weakened to a time integral result\,: 
\begin{theorem}
For any value of $1\leq m < \infty$, if the integral condition is satisfied
\bel{4ththmA}
\int_{0}^{t}\ln \left(\frac{1 + Z_{m}}{c_{4,m}}\right)\,d\tau \geq  0\,,\qquad\qquad Z_{m} = D_{m+1}/D_{m}
\ee
with $c_{4,m} = \left[ 2^{\rho_{m}-1}\left(1 + c_{1,m}c_{2,m}\right)\right]^{ \rho_{m}^{-1}}$, 
then $D_{m}(t) \leq D_{m}(0)$ on the interval $[0,\,t]$. 
\end{theorem}
Given the nature of $c_{4,m}\searrow 2$ it is clear that there must be enough regions of the time axis where 
$D_{m+1} > (c_{4,m}-1)D_{m}$ to make the integral positive. The $D_{m}$ are easily computable from 
Navier-Stokes data. Therefore, an interesting numerical experiment would be to test\,:
\ben
\item Whether the $D_{m}$  are ordered as time evolves such that $D_{m} \geq D_{m+1}$ or  $D_{m} \leq D_{m+1}$?
\item Do the $D_{m}$ cross over from one regime to the other?
\item How significant are the initial conditions and the Reynolds number in this behaviour?
\een

\section{\large Body-forced Navier-Stokes equations}\label{body}

\subsection{A critical lower bound on $\int_{0}^{t}D_{m}\,d\tau$ in terms of $Re$}\label{lb}

In \cite{JDGJMP12,JDGPRS10} the body-forced Navier-Stokes equations were considered in terms of the Grashof number 
$Gr$. It is more useful to to consider this in terms of the Reynolds number $Re$. The inclusion of the forcing in (\ref{Dminequal}) 
modifies\footnote{For the forced case the definition of $\Omega_{m}$ requires an additive $\varpi_{0}$ term 
to act as a lower bound \cite{JDGJMP12,JDGPRS10}.} this but requires the introduction of a third frequency $\varpi_{3,m} = 
\varpi_{0}\alpha_{m}c_{3,m}$ 
\bel{DminequalG}
\dot{D}_{m} \leq D_{m}^{3}\left\{- \varpi_{1,m}\left(\frac{D_{m+1}}{D_{m}}\right)^{\rho_{m}} 
+ \varpi_{2,m}\right\} + \varpi_{3,m}Re^{2} D_{m}\,,
\ee
where $\rho_{m} = \frac{2}{3} m(4m+1)$ and $\gamma_{m} = \frac{1}{2}\alpha_{m+1}\left(m^{2}-1\right)^{-1}$.  
Let $\Delta_{m}$  be defined by $(2 \leq \Delta_{m} \leq 6)$
\bel{5thm2}
\Delta_{m} =  3\left\{\delta_{m}(2+\rho_{m}\gamma_{m}) - \rho_{m}\gamma_{m}\right\}
\ee
The following result shows that if a critical lower bound is set on $ \int_{0}^{t} D_{m}\,d\tau$ then $D_{m}$ 
will decay exponentially. Note that the case $m=1$ is excluded\,:
\par\noindent
\begin{theorem}\label{4ththm}
If there exists a value of $m$ lying in the range $1 < m < \infty$, with 
initial data $\left[D_{m}(0)\right]^{2} < C_{m}Re^{\Delta_{m}}$, for which the integral lies 
on or above the critical value
\bel{5thm1}
c_{m}\left(t\,Re^{3\delta_{m}} + \eta_{2}\right) \leq \int_{0}^{t} D_{m}\,d\tau
\ee
and $\delta_{m}$ and $\eta_{2}$ lie in the ranges 
\bel{deltamdef}
\frac{2/3 + \rho_{m}\gamma_{m}}{2+\rho_{m}\gamma_{m}} < \delta_{m} < 1\,,
\qquad\mbox{and}
\qquad \eta_{2} \geq \eta_{1}Re^{3(\delta_{m}-1)}\,,
\ee
then $D_{m}(t)$ decays exponentially on $[0,\,t]$.
\end{theorem}
\par\medskip\noindent
\textbf{Remark:} $\delta_{m} \searrow 1/2$ for large $m$ so enough slack lies between the upper and lower bounds 
on $\int_{0}^{t}D_{m}\,d\tau$. 
\par\smallskip\noindent
\textbf{Proof\,:} To proceed, divide by $D_{m}^{3}$ to write (\ref{DminequalG}) as
\bel{2ndA}
\frac{1}{2}\frac{d~}{dt}\left(D_{m}^{-2}\right) \geq X_{m}\left(D_{m}^{-2}\right) - \varpi_{2,m}
\qquad\qquad
X_{m} = \varpi_{1,m}\left(\frac{D_{m+1}}{D_{m}}\right)^{\rho_{m}}D_{m}^{2} - \varpi_{3,m}Re^{2}\,.
\ee
A lower bound for $\int_{0}^{t}X_{m}d\tau$ can be estimated thus\,:
\beq{2ndE}
\int_{0}^{t} D_{m+1}\,d\tau 
&=& 
\int_{0}^{t}\left[
\left(\frac{D_{m+1}}{D_{m}}\right)^{\rho_{m}}D_{m}^{2}\right]^{\frac{1}{\rho_{m}}}D_{m}^{\frac{\rho_{m} - 2}{\rho_{m}}}
\,d\tau\non\\ 
&\leq &
\left(\int_{0}^{t}\left(\frac{D_{m+1}}{D_{m}}\right)^{\rho_{m}}D_{m}^{2}\,d\tau\right)^{\frac{1}{\rho_{m}}}
\left(\int_{0}^{t}D_{m}\,d\tau\right)^{\frac{\rho_{m}-2}{\rho_{m}}}t^{1/\rho_{m}}
\eeq
and so
\bel{2ndF}
\int_{0}^{t}X_{m}d\tau \geq \varpi_{1,m}
t^{-1}\frac{\left(\int_{0}^{t}D_{m+1}\,d\tau\right)^{\rho_{m}}}
{\left(\int_{0}^{t}D_{m}\,d\tau\right)^{\rho_{m}-2}}  - \varpi_{3,m}t\,Re^{2}\,.
\ee
Recall that $\rho_{m} = \frac{2}{3} m(4m+1)$ and $\gamma_{m} = \frac{1}{2}\alpha_{m+1}\left(m^{2}-1\right)^{-1}$. 
It is not difficult to prove that $\Omega_{m}^{m^{2}} \leq \Omega_{m+1}^{\,m^{2}-1}\Omega_{1}$
for $m > 1$, from which, after manipulation into the $D_{m}$ becomes
\bel{newDm}
D_{m} \leq D_{m+1}^{\alpha_{m}/2\gamma_{m}m^{2}} D_{1}^{\alpha_{m}/2m^{2}}\qquad\qquad
\frac{\alpha_{m}}{2m^{2}}\left(1+\gamma_{m}^{-1}\right) = 1
\ee
and therefore a H\"older inequality gives
\bel{newthm2}
\frac{\int_{0}^{t}D_{m+1}\,d\tau}{\int_{0}^{t}D_{m}\,d\tau} \geq 
\left(\frac{\int_{0}^{t}D_{m}\,d\tau}{\int_{0}^{t}D_{1}\,d\tau}\right)^{\gamma_{m}}\,.
\ee
\beq{newthm3}
\int_{0}^{t}X_{m}d\tau 
&\geq& \varpi_{1,m}t^{-1} 
\frac{\left(\int_{0}^{t}D_{m}\,d\tau\right)^{\rho_{m}\gamma_{m}+2}}{\left(\int_{0}^{t}D_{1}\,d\tau\right)^{\rho_{m}\gamma_{m}}}
 - \varpi_{3,m}t\,Re^{2}\,.
\eeq
Inequality (\ref{2ndA}) integrates to
\bel{2ndD}
 \left[D_{m}(t)\right]^{2} \leq \frac{\exp\left\{-2\int_{0}^{t}X_{m}\,d\tau\right\}}
{\left[D_{m}(0)\right]^{-2} 
- 2\varpi_{2,m}\int_{0}^{t}\exp\left\{-2\int_{0}^{\tau}X_{m}\,d\tau'\right\}d\tau}\,.
\ee
 (\ref{2ndF}) can be re-written as
\beq{newthm3}
\int_{0}^{t}X_{m}d\tau 
&\geq& \varpi_{1,m}t^{-1} 
\frac{\left(\int_{0}^{t}D_{m}\,d\tau\right)^{\rho_{m}\gamma_{m}+2}}{\left(\int_{0}^{t}D_{1}\,d\tau\right)^{\rho_{m}\gamma_{m}}}
 - \varpi_{3,m}t\,Re^{2}\non\\
&\geq& c_{m}t\left(\varpi_{1,m} Re^{\Delta_{m}} - \varpi_{3,m}Re^{2}\right)
\eeq
having used the assumed lower bound in the theorem and the upper bound of $\int_{0}^{t}D_{1}\,d\tau$.
Moreover, to have the dissipation greater than forcing requires $\Delta_{m} > 2$ so $\delta_{m}$ must lie 
in the range as in (\ref{deltamdef}) because $2 < \Delta_{m} \leq 6$. 
For large $Re$ the negative $Re^{2}$-term in (\ref{2ndD}) is dropped so the integral in 
the denominator of (\ref{2ndD}) is estimated as
\beq{newthm5}
\int_{0}^{t}\exp\left(-2\int_{0}^{\tau}X_{m}d\tau'\right)\,d\tau 
\leq \left[2\tilde{c}_{m}\varpi_{1,m}\right]^{-1}Re^{-\Delta_{m}}
\left(1 - \exp\left[- 2\varpi_{1,m}\tilde{c}_{m}t\, Re^{\Delta_{m}}\right]\right)\,,
\eeq
and so the denominator of (\ref{2ndD}) satisfies
\bel{newthm6}
\mbox{Denominator} \geq \left[D_{m}(0)\right]^{-2} - c_{2,m}c_{1,m}\left(2\tilde{c}_{m}\right)^{-1}
Re^{-\Delta_{m}}\left(1 - \exp\left[- 2\varpi_{1,m}\tilde{c}_{m}t\,Re^{\Delta_{m}}\right]\right)\,.
\ee
This can never go negative if $ \left[D_{m}(0)\right]^{-2} >  
c_{1,m}c_{2,m}\left(2\tilde{c}_{m}\right)^{-1}Re^{-\Delta_{m}}$, 
which means $D_{m}(0) < C_{m}Re^{\shalf\Delta_{m}}$.  \hfill $\blacksquare$

.\par\vspace{-27mm}
\bc
\begin{minipage}[htb]{9cm}
\setlength{\unitlength}{.9cm}
\begin{picture}(11,11)(0,0)
\put(0,0){\vector(0,1){8}}
\multiput(0,0)(0.1,0){100}{.}
\put(-0.75,7.5){\makebox(0,0)[b]{$D_{m}(t)$}}
\put(-0.5,3){\makebox(0,0)[b]{$Re^{3}$}}
\put(2.5,3.1){\makebox(0,0)[b]{\tiny~~~~~~~~~~~~~~~~~~~~~~upper~bound~of~$\left<D_{m}\right>_{t}$}}
\put(3,1.6){\makebox(0,0)[b]{\tiny~~~~~~~~~~~~~~critical~lower-bound~of~$\left<D_{m}\right>_{t}$}}
\put(-.6,1.5){\makebox(0,0)[b]{$Re^{3\delta_{m}}$}}
\put(10.3,-.1){\makebox(0,0)[b]{$t$}}

\multiput(6,0)(0,.1){50}{.}
\multiput(8,0)(0,.1){60}{.}
%
\multiput(0,3)(.1,0){92}{.}
\multiput(0,1.5)(.1,0){92}{.}
\put(0,-.5){\makebox(0,0)[b]{\small $t_{0}$}}
\put(6.1,-.5){\makebox(0,0)[b]{\small $t_{1}$}}
\put(8.1,-.5){\makebox(0,0)[b]{\small $t_{2}$}}
\put(7.1,1){\makebox(0,0)[b]{\tiny potential singularities}}
\put(7.1,.7){\makebox(0,0)[b]{$\uparrow$}}
\put(6.7,.3){\makebox(0,0)[b]{$\uparrow$}}
\qbezier[500](0.01,8)(-.1,-0.25)(6,0.05)
\qbezier[500](6,0.05)(8,0)(8,6)
\end{picture}
\end{minipage}
\ec
\par\vspace{3mm}
\bc
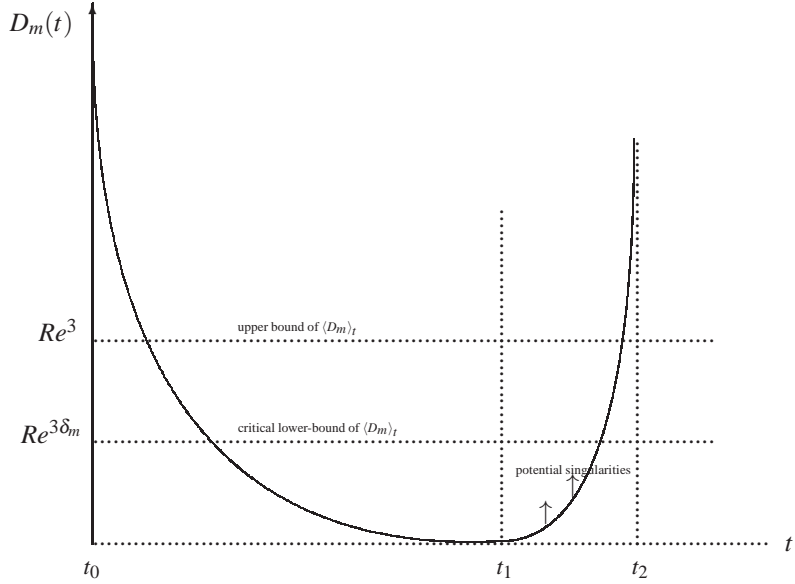
{\textbf{Figure 1\,:} \small A cartoon of $D_{m}(t)$ versus $t$ illustrating the phases of intermittency. 
The range of $\delta_{m}$ lies in Theorem \ref{4ththm}. The vertical arrows depict the region where 
there is the potential for needle-like singular behaviour \& thus a break-down of regularity.}
\ec

\subsection{A relaxation oscillator mechanism for intermittency}\label{inter} 

Experimentally, signals go through cycles of growth and collapse \cite{BT49,KC71,Sreenivasan85,MS91} so it is not realistic to expect 
that the critical lower bound imposed in Theorem \ref{4ththm} should hold for all time. Using the average notation 
$\left<\cdot\right>_{t}$, inequality (\ref{2ndD}) shows that if $\left<D_{m}\right>_{t}$ lies above critical 
then $D_{m}(t)$ collapses exponentially. In Figure 1 the horizontal line at $Re^{3\delta_{m}}$ is drawn 
as the critical lower bound on $\left<D_{m}\right>_{t}$.

Above this critical range, $D_{m}(t)$ will decay exponentially fast. However, because integrals take account of 
history, there will be a delay before $\left<D_{m}\right>_{t}$ decreases below the value above which a zero 
in the denominator of (\ref{2ndD}) can be prevented (at $t_{1}$)\,: at this point all constraints are removed 
and $D_{m}(t)$ is free to grow rapidly again in $t_{1} \leq t \leq t_{2}$.  If the integral drops below critical 
then it is in this interval that the occurrence of singular events (depicted by vertical arrows) must still formally 
be considered -- if one occurs the solutions fails. Provided a solution still exists, growth in $D_{m}$ will be such 
that, after another delay, it will force $\left<D_{m}\right>_{t}$ above critical and the system, with a re-set of initial 
conditions at $t_{2}$, is free to move through another cycle. Thus it behaves like a relaxation oscillator. The vertical 
arrows in Figure 1 label the region, below critical, where
\bel{belowcrit}
\int_{0}^{t} D_{m}\,d\tau < c_{m}\left(t\,Re^{3\delta_{m}} + \eta_{2}\right) \,.
\ee
It is in this regime where where potentially singular point-wise growth of $D_{m}(t)$ could occur which contributes 
little to the growth of the integral $\int_{0}^{t}D_{m}\,d\tau$ and which does not drive it past critical. No control 
mechanism for this type of growth is known and so the regularity problem remains open. 

\section{\large Conclusion}

The variables $D_{m}$, as defined in (\ref{Dmdef}), have proved useful in expressing the Navier-Stokes 
and Euler regularity problems in a natural manner. Their use also poses some interesting questions. For 
instance, while the $\Omega_{m}$ must be ordered because of H\"older's inequality, this is not the case 
with the $D_{m}$ because the $\alpha_{m}$ {\it decrease} with $m$. Theorem \ref{Dmthm} suggests 
that the regime $D_{m+1}/D_{m} \geq c_{m}$, where $c_{m}$ is a constant only just above unity, 
guarantees the decay of $D_{m}$ and hence control over Navier-Stokes solutions. In terms of numerical 
experiments, it would be interesting to see, from a variety of initial conditions, which of the two regimes 
\bel{regimes}
D_{m+1}\geq D_{m} \qquad \mbox{or}\qquad D_{m+1}\leq D_{m}
\ee
are predominant and whether there is a cross-over from one to the other. If so, does this depend heavily 
on the initial conditions, such as the contrasting random or anti-parallel vortex initial conditions? Does it 
also depend on the size of $Re$? Likewise, do solutions of the Euler equations, when in their intermediate 
and late growth phases, track a singular upper bound as in (\ref{test1})?  
\par\smallskip\noindent
\textbf{Acknowledgement\,:} I would like to thank Darryl Holm of Imperial College London and Bob Kerr 
of the University of Warwick for discussions on this topic.


\end{document}